\documentclass[12pt]{iopart}

\usepackage{iopams}

\usepackage{amsthm}

\let\p\partial

\let\ds\displaystyle
\newcommand{\uvar}[0]{u}
\newcommand{\anotherpropcoeffsquared}[0]{K}
\newcommand{\xisquared}[0]{k}

\newcommand{\divt}[0]{\theta}

\newcommand{\intermrelthird}[0]{\sigma}
\newcommand{\arbfuningauss}[0]{\psi}
\newcommand{\firstdet}[0]{\varphi}
\newcommand{\intermrel}[0]{\omega}
\newcommand{\seconddet}[0]{\varrho}
\newcommand{\intermrelsec}[0]{\mu}
\newcommand{\propcoeff}[0]{\lambda}

\newcommand{\pctz}[0]{\kappa}

\newcommand{\mce}[0]{e}
\newcommand{\mcf}[0]{f}
\newcommand{\mcg}[0]{g}

\theoremstyle{plain}
\newtheorem{theorem}{Theorem}[section]

\theoremstyle{definition}

\newcommand{\R}{\mathbb{R}}
\newcommand{\bq}{\mbox{\boldmath$q$}}
\newcommand{\bt}{\mbox{\boldmath$t$}}
\newcommand{\boldr}{\mbox{\boldmath$r$}}

\begin{document}
\title[On ideal fibre-reinforced fluid sheets. Geometry and integrability]{On steady motions of an ideal fibre-reinforced fluid in a curved stratum. Geometry and integrability}

\author{
Dmitry K. Demskoi$^1$ and Wolfgang K. Schief$^2$}
\address{$^1$ School of Computing and Mathematics,
Charles Sturt University, NSW 2678, Australia}

\address{$^2$ School of Mathematics and Statistics,
The University of New South Wales,
Sydney, NSW 2052, Australia}

\eads{\mailto{ddemskoy@csu.edu.au}, \mailto{w.schief@unsw.edu.au}}

\begin{abstract}
It is shown that the kinematic equations governing steady motions of an ideal fibre-reinforced fluid in a curved stratum may be expressed entirely in terms of the intrinsic Gauss equation, which assumes the form of a partial differential equation of third order, for the surface representing the stratum. In particular, the approach adopted here leads to natural non-classical orthogonal coordinate systems on surfaces of constant Gaussian curvature with one family of coordinate lines representing the fibres. Integrable cases are isolated by requiring that the Gauss equation be compatible with another third-order hyperbolic differential equation. In particular, a variant of the integrable Tzitz\'eica equation is derived which encodes orthogonal coordinate systems on pseudospherical surfaces. This third-order equation is related to the Tzitz\'eica equation by an analogue of the Miura transformation for the (modified) Korteweg-de Vries equation. Finally, the formalism developed in this paper is illustrated by focussing on the simplest ``fluid sheets'' of constant Gaussian curvature, namely the plane, sphere and pseudosphere.
\end{abstract}

%
\vspace{2pc}

\section{Introduction}

\newcommand{\eqref}[1]{(\ref{#1})}
\let\cd\cdot

In a series of papers \cite{schiefquart,schieftmf,schiefjmp,murugesh}, the geometry and integrable structure residing in the differential equations governing the motion of ideal fibre-reinforced fluids have been investigated in great detail. Recently, this investigation has culminated in an overarching description of the general non-steady (planar) two-dimensional case which captures and extends the results obtained previously \cite{demschief}. The mathematical theory of the deformation of fibre-reinforced materials has been set down in a monograph by Spencer \cite{spencer}. Hull et al.\ \cite{hull} adopted an ideal fibre-reinforced fluid model to describe resin  matrix fibre-reinforced materials in the important formation state. This model consists of an incompressible fluid which is inextensible along ``fibre'' lines. The latter occupy the volume of the fluid by which they are convected. The presence of these privileged fibre orientations in the fluid imposes strong kinematic constraints on its admissible motions. 

Fully three-dimensional motions of fibre-reinforced fluids have not yet been amenable to the techniques employed in the above-mentioned series of papers. In fact, Spencer \cite{spencer2} has observed that two-dimensional flows of an ideal fibre-reinforced fluid are privileged in that these are essentially determined by kinematic considerations. Therefore, pressure $p$ and tension $T$ in the fibre direction can always be determined such that the equations of motion are satisfied. Against this background, the results presented in this paper have had its origin in the quest to determine to what extent the methods applied in the planar two-dimensional case may be transferred to the case of fibre-reinforced ``fluid sheets'', that is, motions of fibre-reinforced fluids in thin curved strata which are represented by curved surfaces \cite{Weatherburn25}. It turns out that, remarkably, the results of this investigation may not only be of relevance to mathematical physicists who are interested in the mathematical description of such motions but also differential geometers and researchers whose expertise is integrable systems theory.

Here, we show that the kinematic differential equations governing the steady motion of an ideal fibre-reinforced fluid in a curved stratum may be reduced to the third-order partial differential equation 
\begin{equation}\label{key}
	\left(\frac{u_{ns}}{\arbfuningauss}\right)_n + (\arbfuningauss''+\anotherpropcoeffsquared\arbfuningauss)u_s = 0
\end{equation}
for a function $u$, wherein $\psi=\psi(u)$ is an arbitrary function depending on $u$ only. In fact, $u$ represents arc length along the fibre lines on the surface $M^2$ on which the motion takes place. Here, $K$ denotes the Gaussian curvature of $M^2$. The metric of $M^2$ is entirely parametrised in terms of $u$ and $\psi$ if one chooses the fibre lines ($s$-lines) and their orthogonal trajectories ($n$-lines) as the coordinate lines on $M^2$ so that the above third-order differential equation is nothing but the Gauss equation~\cite{Eisenhart60} associated with the metric. The geometric nature of the kinematic equations of fibre-reinforced fluids is, in fact, already encoded in their original formulation since, in purely mathematical terms, these capture the commutativity of two tangent vector fields with one of them being divergence free and the other one being normalised to unity.

It turns out that large classes of motions on surfaces $M^2$ of constant Gaussian curvature $K$ are governed by a Gauss equation which is integrable in the sense of soliton theory \cite{AblowitzClarkson1991,rogersschief}.  In particular, the integrable structure discovered in \cite{schiefquart,schieftmf} in the planar steady case may be retrieved in the current setting by considering the case $K=0$. These integrable cases are isolated by exploiting an observation made by Adler and Shabat \cite{adler} who have noticed that integrable hyperbolic equations of third-order frequently admit compatible third-order differential equations. Even though the existence of a compatible equation does not necessarily imply integrability, we prove that, in the current context,  each constraint on the arbitrary function $\psi(u)$ generated by this approach renders the Gauss equation (\ref{key}) integrable. In the simplest case, the Gauss equation may be integrated once to obtain the sin(h)-Gordon equation
\begin{equation}
  u_{ns} = \sin u\quad(K>0),\qquad u_{ns} = \sinh u\quad(K<0).
\end{equation}
It is important to point out that the classical theory of surfaces of constant Gaussian curvature $K=\pm 1$ identifies a sin(h)-Gordon equation of the form \cite{rogersschief}
\begin{equation}
  \omega_{xx} + \omega_{yy} + \sinh\omega = 0\quad(K>0),\qquad \omega_{xx} - \omega_{yy} = \sin\omega\quad (K<0)
\end{equation}
as the Gauss equation, where $x$ and $y$ constitute curvature coordinates \cite{Eisenhart60}. Hence, as a by-product of the current investigation, it has been established that there exist natural non-classical orthogonal coordinate systems on surfaces of constant Gaussian curvature.

Remarkably, there exists another class of ``integrable'' motions of fibre-reinforced fluids on surfaces of constant negative Gaussian curvature which is related to an integrable differential equation arising in an entirely different classical geometric context. Thus, the discovery by Tzitz\'eica \cite{Tzitzeica0708,rogersschief} of a class of surfaces, the properties of which are preserved by affine transformations, is widely regarded as the beginning of the theory of affine differential geometry. Tzitz\'eica demonstrated that these ``affine spheres'' are governed by the partial differential equation
\begin{equation}\label{tzitzeica}
  \varphi_{xy} = e^\varphi - e^{-2\varphi},
\end{equation}
where $x$ and $y$ constitute asymptotic coordinates \cite{Eisenhart60}. We demonstrate that the Gauss equation descriptive of the afore-mentioned class of motions is related to the Tzitz\'eica equation (\ref{tzitzeica}) by an analogue of the celebrated Miura transformation \cite{Miura1968} connecting the Korteweg-de Vries (KdV) equation and its modified analogue, the modified KdV (mKdV) equation \cite{AblowitzClarkson1991}. Accordingly, the ``modified Tzitz\'eica equation'' obtained here gives rise to a canonical orthogonal coordinate system on surfaces of constant negative Gaussian curvature. In this connection, it is observed that a systematic examination of coordinate systems on such pseudospherical surfaces which are defined in terms of integrable systems may be found in \cite{Sasaki79,ChernTenenblatt1986}.

Due to the intrinsic nature of the Gauss equation, the determination of the associated motions on any concrete surface $M^2$ embedded in $\mathbb{R}^3$ requires the derivation of the Frobenius system which encapsulates the coordinate transformation between the adapted orthogonal coordinates $(s,n)$ and the coordinates which parametrise the given surface $M^2$. Here, we solve this ``embedding problem'' for the simplest surfaces of curvature $K=0$ and $K=\pm 1$, namely, the plane, sphere and pseudosphere. It turns out that the corresponding coordinate transformations are given in terms of quadratures and linear Frobenius systems which are compatible modulo the Gauss equation. 

\section{The governing equations}
\label{eqs}

\subsection{The mathematical setup}

Mathematically, the steady motion of a fibre-reinforced fluid in a curved stratum may be described in the following manner \cite{spencer,Weatherburn25}. The shape of the fluid stratum is encapsulated in a surface $M^2$ embedded in Euclidean space $\R^3$ on which the motion takes place. Let
\begin{equation}
g = g_{ij} dx^idx^j
\end{equation}
be the metric tensor \cite{Eisenhart49} on $M^2$ in terms of local coordinates $x^1$ and $x^2$. Here, the usual rule of summation over repeated indices is adopted. The motion is governed by a system of differential equations for two vector fields $\bq$ and $\bt$ on $M^2$. In tensor notation, the vector field $q^i$ represents the velocity of the fluid, while $t^i$ is a unit vector field, that is,
\begin{equation}\label{t1}
 \|{\bt}\|^2=g_{ij}t^i t^j=1,
\end{equation}
the integral curves of which represent the fibres embedded in the fluid. The fluid velocity obeys the continuity equation which, in the current context, reduces to the vanishing divergence condition
\begin{equation}
 \mbox{div}\, {\bq}=\nabla_i q^i=0
	\label{divq0}
\end{equation}
with $\nabla_i$ denoting the covariant derivative with respect to $x^i$. Finally, the requirement that the fibres be convected with the fluid translates into the commutativity condition
\begin{equation}
	q^i \nabla_i t^j-t^i \nabla_i q^j=0.
	\label{eq0}
\end{equation}
In summary, the governing equations for the steady motion of a fibre-reinforced fluid are given by the differential equations (\ref{divq0}) and (\ref{eq0}) subject to the algebraic constraint (\ref{t1}).

It turns out that the fibre divergence
\begin{equation}
	\divt=\mbox{div}\, \bt=\nabla_i t^i
	\label{divt}
\end{equation}
constitutes an important quantity in the subsequent discussion. Indeed, if we calculate the divergence of (\ref{eq0}) then we obtain
\begin{equation}
\eqalign{
0 = \nabla_j(q^i \nabla_i t^j-t^i \nabla_i q^j)\\ \phantom{0} =\left(\nabla_{j} q^{i}\right) \nabla_{i} t^{j} 
+q^{i} \nabla_{j} \nabla_{i} t^{j}-\left(\nabla_{j} t^{i}\right) \nabla_{i} q^{j}-t^{i} \nabla_{j} \nabla_{i} q^{j}.}\end{equation}
Application of the standard identities \cite{Eisenhart49}
\begin{equation}
\eqalign{
\nabla_{j} \nabla_{i} t^{j} &= \nabla_{i} \nabla_{j} t^{j}-R_{k i} t^k\\
\nabla_{j} \nabla_{i} q^{j} &=-R_{k i} q^k,}
\end{equation}
where $R_{ik}=R_{ki}$ is the symmetric Ricci tensor, leads to
\begin{equation}
0 = q^i \nabla_i	\nabla_j t^j=q^i \nabla_i \divt
\label{convtheta}
\end{equation}
so that the fibre divergence is convected with the fluid. 

\subsection{An adapted orthogonal coordinate system}

We now choose a coordinate system which is adapted to the geometry of the fibres. Thus, if $x^1 = s$ and $x^2 = n$ parametrise the fibre lines and their orthogonal trajectories respectively then the metric on $M^2$ adopts the diagonal form
\begin{equation}
	g = \alpha^2 ds^2 + \beta^2 dn^2
	\label{2dmetric}
\end{equation}
and the components of the unit vector $\bt$ are given by
\begin{equation}
	t^1=\frac{1}{\alpha},\quad t^2=0.
	\label{eqfortn}
\end{equation}
On use of the general expression \cite{Eisenhart49}
\begin{equation}\label{divergence}
  \mbox{div}\,\mbox{\boldmath$w$} = \frac{1}{\sqrt{|g|}}\p_i(\sqrt{|g|}w^i)
\end{equation}
for the divergence of a vector field $\mbox{\boldmath$w$}$, where $|g|=\alpha^2\beta^2$ is the determinant of the metric tensor $g$, the divergence condition (\ref{divq0}) takes the form
\begin{equation}\label{divq}
{(\alpha\beta q^1)}_s+{(\alpha\beta q^2)}_n=0
\end{equation}
with subscripts denoting partial derivatives. We also observe that (\ref{divergence}) applied to the vector field $\bt$ leads to
\begin{equation}
	\divt = \frac{1}{\alpha}\p_s \ln \beta.
	\label{thetaalphabeta}
\end{equation}

The commutativity condition (\ref{eq0}) guarantees the existence of a function $u$ defined by the pair of equations
\begin{equation}
	t^i\nabla_i \uvar=1,\quad q^i \nabla_i \uvar=0.
	\label{eqforrho}
\end{equation}
Thus, $\uvar$ represents arc length along the fibres which is convected with the fluid. Since $u$ cannot be constant, any other function which is convected with the fluid and $u$ must be functionally dependent so that (\ref{convtheta}) implies that
\begin{equation}
\theta = \theta(u).
\end{equation}
Accordingly, the arc length condition (\ref{eqforrho})$_1$ becomes
\begin{equation}
	\alpha=\uvar_s
	\label{rhosalpha}
\end{equation}
and (\ref{thetaalphabeta}) suggests introducing the parametrisation
\begin{equation}
\divt=\frac{\arbfuningauss'(\uvar)}{\arbfuningauss(\uvar)}
\end{equation} 
so that integration results in
\begin{equation}
	\beta=N(n)\arbfuningauss(\uvar),
	\label{betapsi}
\end{equation}
wherein the function of integration $N(n)$ may be scaled to unity if one re-parametrises the 
$n$-lines appropriately. Here, the prime denotes differentiation with respect to $u$. Thus, the metric on $M^2$ adopts the form
\begin{equation}
g = u_s^2ds^2 + \psi^2dn^2.
\end{equation}

\subsection{The commutativity condition}

In order to evaluate the remaining constraint on the vector fields $\bq$ and $\bt$, that is, the commutativity condition (\ref{eq0}), it is recalled that the covariant derivative of a vector field is given by \cite{Eisenhart49}
\begin{equation}
 \nabla_i w^j = \p_i w^j + {\Gamma^j}_{ik}w^k,
\end{equation}
wherein the Christoffel symbols are defined by
\begin{equation}
{\Gamma^j}_{ik} = \frac{1}{2}g^{jm}(\p_kg_{mi} + \p_ig_{mk} - \p_mg_{ik}),
\end{equation}
with $g^{jm}$ being the inverse of the metric tensor. Since the only non-vanishing components of the metric tensor and its inverse are given by
\begin{equation}
 g_{11} = u_s^2,\quad g_{22} = \psi^2,\quad g^{11}=u_s^{-2},\quad g^{22}=\psi^{-2},
\end{equation}
the Christoffel symbols are calculated to be
\begin{equation}
\eqalign{
	{\Gamma^1}_{11}=\frac{\uvar_{ss}}{\uvar_s},\quad {\Gamma^1}_{12}=\frac{\uvar_{sn}}{\uvar_s},\quad {\Gamma^1}_{22}=-\frac{\arbfuningauss \arbfuningauss'}{\uvar_s}\\
	{\Gamma^2}_{11}=-\frac{\uvar_s\uvar_{sn}}{\arbfuningauss^2},\quad {\Gamma^2}_{12}=\frac{\arbfuningauss'\uvar_s}{\arbfuningauss},\quad {\Gamma^2}_{22}=\frac{\arbfuningauss'\uvar_n}{\arbfuningauss}.}
	\label{gammas}
\end{equation}
Hence, we obtain
\begin{equation}
	\nabla_1 t^1=0,\quad \nabla_2 t^1=0,\quad
	\nabla_1 t^2={\Gamma^2}_{11}t^1, \quad \nabla_2 t^2={\Gamma^2}_{12}t^1.
	\label{covdivs}
\end{equation}

Now, the commutativity condition (\ref{eq0})$_{j=2}$ reduces to
\begin{equation}
q^2_s = 0,
\end{equation}
while (\ref{eq0})$_{j=1}$ may be formulated as
\begin{equation}
{(u_sq^1)}_s + u_{sn}q^2 = 0.
\end{equation}
The latter is a differential consequence of the convection condition (\ref{eqforrho})$_1$ given by
\begin{equation}\label{conv}
 u_sq^1 + u_n q^2 = 0.
\end{equation}
Finally, elimination of $q^1$ between (\ref{conv}) and the divergence condition (\ref{divq}), which reads
\begin{equation}
{(u_s\psi q^1)}_s + {(u_s\psi q^2)}_n = 0,
\end{equation}
results in
\begin{equation}
q^2_n = 0.
\end{equation}
We therefore conclude that
\begin{equation}
 q^1 = -c\frac{u_n}{u_s},\quad q^2 = c = \mbox{const}.
\end{equation}
It is observed that the constant $c$ reflects the fact that the original system of governing equations is linear and homogenous in the fluid velocity $\bq$ so that it merely corresponds to a scaling of the fluid flow.

\subsection{Motions on surfaces of constant Gaussian curvature}

Motions on surfaces of constant Gaussian curvature $K$ naturally generalise planar motions for which  $K=0$. The geometry and integrability of the latter have been discussed in detail in \cite{schieftmf,schiefjmp}. In general, the Gaussian curvature is related to the metric coefficients by Gauss' {\em Theorema Egregium} which is given by \cite{Eisenhart60}
\begin{equation}
\left(\frac{\alpha_n}{\beta}\right)_n + \left(\frac{\beta_s}{\alpha}\right)_s + \anotherpropcoeffsquared \alpha \beta = 0
	\label{gauss}
\end{equation}
in the case of a metric of the form (\ref{2dmetric}). The preceding analysis with $\alpha=u_s$ and $\beta=\psi(u)$ therefore gives rise to the main result of this section.

\begin{theorem} 
The $s$-lines of an orthogonal coordinate system $(x^1,x^2) = (s,n)$ on a surface $M^2$ of constant Gaussian curvature $K$ may be regarded as the fibre lines in a steady motion of a fibre-reinforced fluid if and only if the metric on $M^2$ is of the form
\begin{equation}\label{metric}
  g = u_s^2 ds^2+\psi^2 dn^2,
\end{equation}
where $\psi(u)$ is an arbitrary function of $u$ only, and $u$ obeys the nonlinear third-order differential equation
\begin{equation}
	\left(\frac{u_{ns}}{\arbfuningauss}\right)_n + (\arbfuningauss''+\anotherpropcoeffsquared\arbfuningauss)u_s = 0.
	\label{mtzgen}
\end{equation}
The velocity of the fluid and the unit tangent to the fibres are given respectively by
\begin{equation}
 \bq =c\left(-\frac{u_n}{u_s},1\right),\quad \bt=\left(\frac{1}{u_s},0\right),
\end{equation}
where $c$ is an arbitrary constant. Any generic motion of a fibre-reinforced fluid on a surface of constant Gaussian curvature is captured in this manner.
\end{theorem}

\section{Isolation of integrable cases via compatible constraints}

In general, finding non-trivial classes of solutions of the nonlinear governing equation (\ref{mtzgen}) constitutes a highly non-trivial task. However, it turns out that the case of constant Gaussian curvature
as considered in the above theorem is privileged. A first confirmation of this assertion is the observation that if we demand that the function $\psi$ be a solution of the second-order equation $\psi'' + K \psi = 0$ then the third-order equation may be integrated to obtain essentially
\begin{equation}\label{linearisable}
  u_{sn} = G(u),\quad G(u)\in\{0, u, \exp u,\sin u, \sinh u, \cosh u\},
\end{equation}
depending on the sign of $K$. Accordingly, the linear wave and Klein-Gordon equations, the explicitly solvable Liouville equation and the integrable sine-, sinh-, and cosh-Gordon equations \cite{AblowitzClarkson1991,rogersschief} govern particular classes of motions on surfaces of constant Gaussian curvature. This observation is the motivation for the classification scheme presented in this section. Thus, the governing equation (\ref{mtzgen}) is a particular case of a third-order hyperbolic equation and, even though the class of third-order hyperbolic equations is not well studied (in comparison to second-order equations), it is known that it contains many integrable examples (see, e.g., \cite{ff}-\cite{degashone}). In fact, Adler and Shabat \cite{adler} pointed out that an {\em integrable} hyperbolic equation of third order often admits a {\em compatible} equation of the form $u_{nss}=\hat{F}$ with $\hat{F}$ depending on lower-order derivatives of $u$. Hence, even though, in general, compatibility does not imply integrability, we demonstrate below that, in the current setting, this observation is very practical in that it results in all specialisations of  (\ref{mtzgen}) found in this manner being integrable.

It is easily verified that if we assume that (\ref{mtzgen}) admits a compatible equation of the form $u_{nss} = \hat{F}$, where $\hat{F}$ may depend on all derivatives of $u$ up to second order, then the associated compatibility condition immediately implies that $\hat{F}$ cannot depend on $u_n$ and $u_{nn}$. Thus, for convenience, we may bring the compatible equation to be considered into the form
\begin{equation}
	\left(\frac{u_{ns}}{\arbfuningauss}\right)_s=F(u,u_s,u_{ss},u_{ns}).
	\label{compateq}
\end{equation}
The compatibility condition 
\begin{equation}
	\left(\frac{u_{ns}}{\arbfuningauss}\right)_{ns}=\left(\frac{u_{ns}}{\arbfuningauss}\right)_{sn}
	\label{compatty}
\end{equation}
of (\ref{mtzgen}) and (\ref{compateq}) produces the overdetermined system 
\begin{equation}
\eqalign{
	\frac{\p F}{\p u_s}u_{ns}
	=-\frac{\p F}{\p u_{ss}}\left(u_{ns} u_s\frac{\arbfuningauss'}{\arbfuningauss}+\arbfuningauss F\right)\\
    \phantom{\frac{\p F}{\p u_s}u_{ns}}+\left(\frac{\p F}{\p u_{ns}}u_s-\frac{u_{ss}}{\arbfuningauss}\right)(\firstdet+\anotherpropcoeffsquared\arbfuningauss^2+\arbfuningauss'^2)\\
	\phantom{\frac{\p F}{\p u_s}u_{ns}}-\frac{u_s^2}{\arbfuningauss^2}[\arbfuningauss'^3+(\firstdet\arbfuningauss)'+\anotherpropcoeffsquared\arbfuningauss^2\arbfuningauss'],\\
	\frac{\p F}{\p u}=-u_{ns}\frac{\arbfuningauss'}{\arbfuningauss}\frac{\p F }{\p u_{ns}}}
	\label{twocomp}
\end{equation}
for the function $F$, where
\begin{equation}\label{condphi}
\firstdet=\arbfuningauss''\arbfuningauss-\arbfuningauss'^2.
\end{equation}
Here, we exploit the fact that $F$ does not depend on $u_n$, which allows us to split (\ref{compatty}) into the pair of conditions $(\ref{twocomp})_1$ and $(\ref{twocomp})_2$.
The second equation is easy to integrate, but we postpone this until we go through the compatibility conditions which will enable us to determine the function $\arbfuningauss$. Thus, the compatibility condition may be written as
\begin{equation}
	\eqalign{
0&=\frac{\p^2 F}{\p u_s\p u}-\frac{\p^2 F}{\p u\p u_s}\cr
&= \frac{u_s}{u_{ns}}\left[\frac{\firstdet u_{ns}}{\arbfuningauss^2}\frac{\p F}{\p u_{ss}}-\firstdet' \frac{\p F}{\p u_{ns}}
+ \frac{u_{ss}\firstdet'}{u_s \arbfuningauss }
+\frac{u_s}{\arbfuningauss^3}\big(\intermrel-\arbfuningauss\arbfuningauss'\firstdet'+\anotherpropcoeffsquared\firstdet\arbfuningauss^2\big)\right],
}
\label{whenrho}
\end{equation}
where
\begin{equation}
\intermrel=\arbfuningauss^2\firstdet''+\firstdet \arbfuningauss'^2+\firstdet^2.
\end{equation}
It is then natural to examine the following three cases.
\medskip

\noindent
(1) The compatibility condition (\ref{whenrho}) is evidently satisfied if $\firstdet=0,$ in which case the function $\psi$ is determined by
\begin{equation}
	\arbfuningauss'=\varkappa\arbfuningauss.
\end{equation}
\medskip

\noindent
(2) If we assume that $\firstdet'=0$, but $\firstdet\ne 0$, then (\ref{condphi}) is equivalent to
\begin{equation}
\arbfuningauss''=\xisquared\arbfuningauss.
\end{equation}
In this case, we can solve (\ref{whenrho}) for $\p F/\p u_{ss}$ to obtain
\begin{equation}
	\frac{\p F}{\p u_{ss}}=-\frac{u_s}{\arbfuningauss u_{ns}}(\arbfuningauss'^2+\anotherpropcoeffsquared\arbfuningauss^2+\firstdet).
	\label{F_uss}
\end{equation}
One may now verify that the system (\ref{twocomp}), (\ref{F_uss}) is compatible without any further constraints.
\medskip

\noindent
(3) In the case $\firstdet'\ne 0$, the compatibility condition (\ref{whenrho}) may be solved for $\p F/\p u_{ns}$ so that it is required to examine two additional compatibility conditions. One of these conditions is given by
\begin{equation}	
\eqalign{
	0&=\frac{\p^2 F}{\p u_{ns}\p u}-\frac{\p^2 F}{\p u\p u_{ns}}\cr
	&=\frac{\seconddet u_{ns}}{\firstdet'^2\arbfuningauss^2} \frac{\p F}{\p u_{ss}}
	+\frac{u_s}{\arbfuningauss^3\firstdet\firstdet'^2}\left[ \seconddet\intermrel-\arbfuningauss^2(\seconddet'\firstdet'-\anotherpropcoeffsquared\firstdet\seconddet)\right],
}\label{relcom0}
\end{equation}
where
\begin{equation}
\seconddet=\firstdet''\firstdet-\firstdet'^2.
\end{equation}
If we assume that $\seconddet\ne 0$ then (\ref{relcom0}) determines $\p F/\p u_{ss}$ which is required to be compatible with the expression for $\p F/\p u_{ns}$. It turns out that
\begin{equation}
	\frac{\p^2 F}{\p u_{ns}\p u_{ss}}-\frac{\p^2 F}{\p u_{ss}\p u_{ns}}=\frac{u_s}{u_{ns}^2 \arbfuningauss\firstdet} \left( \intermrel+\anotherpropcoeffsquared \arbfuningauss^2\firstdet-\frac{\arbfuningauss^2\seconddet'\firstdet'}{\seconddet}\right)-\frac{1}{u_s\arbfuningauss}
	\label{relcom}
\end{equation}
which cannot vanish for any $\psi$. Hence, $\seconddet=0$ so that
\begin{equation}
	\firstdet''=\frac{\firstdet'^2}{\firstdet}
	\label{seconddet0}
\end{equation}
and (\ref{relcom0}) is identically satisfied. 

The second compatibility condition is given by
\begin{equation}
\eqalign{
 0 = \ds \frac{\p^2 F}{\p u_{s}\p u_{ns}}-\frac{\p^2 F}{\p u_{ns}\p u_{s}}\\
\phantom{0}=\frac{\intermrelsec}{u_{ns}}\frac{\p F}{\p u_{ss}}+\frac{F}{u_s u_{ns}}
-\frac{u_{ss}}{\arbfuningauss u_s^2 }+\frac{u_s^2}{\firstdet\firstdet'\arbfuningauss^3 u_{ns}^2}\intermrelthird_{+}\intermrelthird_{-}+\frac{\firstdet'}{\firstdet\arbfuningauss},}
	\label{compatusuns}
\end{equation}
where
\begin{equation}
\eqalign{
	\ds \intermrelsec=\frac{u_{ss}}{u_{s}}-\frac{\arbfuningauss' u_s}{\arbfuningauss } +\frac{u_s(\intermrel+\anotherpropcoeffsquared \firstdet\arbfuningauss^2)}{\arbfuningauss^2 \firstdet'}-\frac{\arbfuningauss}{u_{ns}} F,\\
	\ds \intermrelthird_{\pm}=\arbfuningauss \arbfuningauss' \firstdet'-\firstdet \arbfuningauss'^2-\firstdet^2-\anotherpropcoeffsquared\firstdet\arbfuningauss^2\pm\sqrt{-\anotherpropcoeffsquared} \firstdet'\arbfuningauss^2.}
	\label{muF}
\end{equation}
It is noted that, {\em a priori}, $K$ could be positive so that $\sigma_{+}$ and $\sigma_{-}$ are complex conjugates. However, the analysis presented below demonstrates that the reality of the function $\psi$ excludes this possibility. Now, once again, the multiplier $\mu$ in (\ref{compatusuns}) requires the distinction between $\mu=0$ and $\mu\neq0$. Thus, if $\mu=0$ then one can verify that the expression for $F$ provided by $(\ref{muF})_1$ does not satisfy (\ref{compatusuns}). Therefore, we now assume that $\intermrelsec\ne 0$ and may solve  (\ref{compatusuns}) for $\p F/\p u_{ss}$. The latter is required to be compatible with $\p F/\p u_{ns}$ obtained from (\ref{whenrho}), leading to
\begin{equation}
\eqalign{
	0 = \ds \frac{\p^2 F}{\p u_{ns}\p u_{ss}}-\frac{\p^2 F}{\p u_{ss}\p u_{ns}}\\
\phantom{0} =\frac{1}{\arbfuningauss^5\firstdet'^2\intermrelsec^2}\left[\left(\frac12\firstdet(\intermrelthird_{+}+\intermrelthird_{-})-2\arbfuningauss^2\firstdet'^2\right)\left(\frac{u_{ns}^2}{u_s}+2\frac{\arbfuningauss^2\firstdet'\intermrelsec}{\firstdet}\right)-u_s\intermrelthird_{+}\intermrelthird_{-}\right].}
	\label{compatunsuss}
\end{equation}

Since the right-hand side of the above compatibility condition contains the function $F$ via the expression $(\ref{muF})_1$ for $\mu$, there exist two possibilities. We first consider the case of not being able to solve (\ref{compatunsuss}) for $F$ so that 
\begin{equation}
	\intermrelthird_{+}\intermrelthird_{-}=0, \quad \firstdet(\intermrelthird_{+}+\intermrelthird_{-})-4\arbfuningauss^2\firstdet'^2=0.
	\label{conds}
\end{equation}
If we integrate (\ref{seconddet0}) to obtain
\begin{equation}
\firstdet'=\propcoeff\firstdet,
\end{equation}
where $\propcoeff$ is a constant of integration, then we can re-write (\ref{muF})$_2$ as
\begin{equation}
\intermrelthird_{\pm}=\firstdet\arbfuningauss(\propcoeff\arbfuningauss'-\arbfuningauss''-\anotherpropcoeffsquared\arbfuningauss\pm\sqrt{-\anotherpropcoeffsquared}\propcoeff\arbfuningauss)
\label{sigmapm}
\end{equation}
and, hence, condition $(\ref{conds})_2$ becomes
\begin{equation}
	\propcoeff \arbfuningauss'-\arbfuningauss''-\anotherpropcoeffsquared\arbfuningauss-2\propcoeff^2\arbfuningauss=0.
	\label{conds2bec}
\end{equation}
Comparison with (\ref{sigmapm}) shows that, in order to satisfy (\ref{conds2bec}) and $(\ref{conds})_1$, we must set
\begin{equation}
\propcoeff=\mp\frac{\sqrt{-\anotherpropcoeffsquared}}{2}
\end{equation}
so that (\ref{conds2bec}) becomes
\begin{equation}
		\arbfuningauss''=\mp\frac{\sqrt{-\anotherpropcoeffsquared}}{2}(\arbfuningauss'\mp\sqrt{-\anotherpropcoeffsquared}\arbfuningauss)
			\label{constraint}
\end{equation}
and it has been verified that $K\leq0$ since $\psi$ is required to be real. We therefore conclude that the function $\psi$ is constrained by the differential equation 
\begin{equation}
	\arbfuningauss''=-\frac{\pctz}{2}(\arbfuningauss'-\pctz\arbfuningauss),
	\label{mtzconstraint}
\end{equation}
where
\begin{equation}
\pctz=\pm\sqrt{-\anotherpropcoeffsquared}.
\end{equation}

In the other case, that is, when (\ref{conds}) is not satisfied, we may solve (\ref{compatunsuss}) for $F$ to obtain
\begin{equation}
\eqalign{
	F=\frac{\firstdet}{2\firstdet'\arbfuningauss^3}\frac{u_{ns}^3}{u_s}+\frac{u_{ns} u_{ss} }{\arbfuningauss u_s}\\
\phantom{F}+u_{ns}u_s\left(
	\frac{\intermrel+\anotherpropcoeffsquared\firstdet\arbfuningauss^2}{\arbfuningauss^3\firstdet'}-\frac{\firstdet\intermrelthird_{+}\intermrelthird_{-}}{\firstdet'\arbfuningauss^3[ \firstdet(\intermrelthird_{+}+\intermrelthird_{-})-4\arbfuningauss^2\firstdet'^2]}-\frac{\arbfuningauss'}{\arbfuningauss^2}
	\right).}
	\label{wrongF}
\end{equation}
However, one may verify that this expression makes the system of equations for $F$ inconsistent. For instance, if (\ref{wrongF}) is substituted into $(\ref{twocomp})_1$ then the terms containing various powers of $u_{ns}$  may not be cancelled regardless of the choice of $\arbfuningauss$. Accordingly, the above analysis may be summarised as follows.

\begin{theorem}
 The Gauss equation (\ref{mtzgen}) admits a compatible equation of the form (\ref{compateq}) if the function $\psi$ satisfies one of the following constraints.
\begin{equation}
\eqalign{
	 (1)\quad\arbfuningauss'=\varkappa\arbfuningauss\quad\Rightarrow\quad  \psi=a e^{\varkappa\uvar}  \\
	 (2)\quad\arbfuningauss'' =\xisquared\arbfuningauss\quad\Rightarrow\quad \psi=a e^{\sqrt{\xisquared} \uvar}+b e^{-\sqrt{\xisquared} \uvar} \\
	 (3)\quad\arbfuningauss'' =-\frac{\pctz}{2}(\arbfuningauss'-\pctz\arbfuningauss)\quad\Rightarrow\quad  \psi=a e^{\pctz \uvar/2}+b e^{-\pctz \uvar}}
	\label{constraints}
\end{equation}
Here, the constants $a,b$ and $\varkappa,k,\kappa$ are arbitrary. In Case (2), $a$ and $b$ are complex conjugates if $k<0$ and, in Case (3), $K = -\kappa^2<0$.
\end{theorem}

\section{Integrability and symmetries}

We now address the significance and, in fact, the applicability of the analysis carried out in the previous section. 

\subsection{Sine-Gordon and modified KdV connections}

We begin by assuming that, as in Case (2),
\begin{equation}
 \psi'' = k\psi,
\end{equation}
which, in fact, includes Case (1) with $k = \varkappa^2$. The Gauss equation (\ref{mtzgen}) may then be formulated as the coupled system of equations
\begin{equation}
  u_{ns} = v\psi,\quad v_n + (K + k)\psi u_s = 0,
\end{equation}
leading to the first integral
\begin{equation}
  (K + k)u_s^2 + v^2 = f(s).
\end{equation}
If $k = -K$ then one reproduces the initial observation made at the beginning of Section~3, leading to the linearisable/solvable/integrable cases (\ref{linearisable}). If $K + k\neq 0$ then one has to distinguish between two cases. Here, we focus on the case $K + k > 0$. The case $K + k <0$ may be dealt with in an analogous manner. Thus, the parametrisation 
\begin{equation}
  \sqrt{K+k}\,u_s = S(s)\sin\rho,\quad v = S(s)\cos\rho,\quad f(s) = [S(s)]^2
\end{equation}
of the first integral results in the coupled system
\begin{equation}
  u_s = \frac{S(s)}{\sqrt{K+k}}\sin\rho,\quad \rho_n = \sqrt{K+k}\,\psi.
\end{equation}
It is noted that an appropriate reparametrisation of the $s$-lines and scaling of $\psi$ may be used to remove the function $S$ and the constant $\sqrt{K+k}$ so that, without loss of generality, we proceed with $S=1$ and $K + k = 1$ to obtain the canonical form
\begin{equation}\label{normalform}
  u_s = \sin\rho,\quad \rho_n = \psi
\end{equation}
which now naturally leads to the following cases.
\medskip

\noindent
(A) If $k=0$ then either $\psi = \mbox{const}$, in which case (\ref{normalform}) may be trivially integrated, or $\psi=u$ without loss of generality so that (\ref{normalform}) becomes the sine-Gordon equation
\begin{equation}\label{sine-Gordon}
  \rho_{sn} = \sin\rho.
\end{equation}
The latter admits large classes of explicit solutions which may be obtained by means of the powerful techniques of integrable systems theory \cite{rogersschief}.
\medskip

\noindent
(B) If $\psi' = \varkappa\psi$ for non-vanishing $\varkappa$ then we may set $\varkappa=1$ and $\psi=e^u$ for illustrative purposes. In this case, (\ref{normalform}) becomes
\begin{equation}
  \rho_{sn} = \rho_n \sin\rho,
\end{equation}
which may be integrated to obtain
\begin{equation}
  \rho_s + \cos\rho = U(s).
\end{equation}
The substitution
\begin{equation}
 \rho = 2\arctan\chi - \frac{\pi}{2}
\end{equation}
then leads to the Riccati equation
\begin{equation}
  \chi_s = \frac{U}{2}(1+\chi^2) -\chi
\end{equation} 
which is well-known to be linearisable. Indeed, the linearising transformation is given by
\begin{equation}
\chi = \frac{y_2}{y_1},
\end{equation}
where $y_1$ and $y_2$ obey the linear system
\begin{equation}\label{scattering}
 \left(\begin{array}{c}y_1\\ y_2\end{array}\right)_s = \frac{1}{2}\left(\begin{array}{cc} 1\, & -U\\ U\, & -1\end{array}\right)\left(\begin{array}{c}y_1\\ y_2\end{array}\right).
\end{equation}
The latter constitutes the scattering problem \cite{AblowitzClarkson1991} for the modified Korteweg-de Vries (mKdV) equation, wherein the spectral parameter is $1$.
\medskip

\noindent
(C) If $\psi'' = k\psi$ for non-vanishing $k$ and $\psi'\not\sim\psi$, we may choose $k = -1$ and $\psi = \sin u$ for simplicity. Other choices of $k$ and $\psi$ may be treated in a similar manner. In the current situation, system (\ref{normalform}) reads
\begin{equation}\label{backlund}
 u_s = \sin\rho,\quad \rho_n = \sin u.
\end{equation}
The latter constitutes a well-known pair of integrable equations, namely the classical B\"acklund equations for the sine-Gordon equation \cite{rogersschief} (with the B\"acklund parameter being~$1$). In fact, it is easy to see that this system implies that
\begin{equation}\label{sine}
  {(u\pm\rho)}_{sn} = \sin(u\pm\rho),
\end{equation}
which indeed shows that the system (\ref{backlund}) provides a link between the two solutions $u\pm\rho$ of the sine-Gordon equation
(\ref{sine}).
\medskip

The integrability of Cases (A) and (C) may also be revealed in an alternative manner. Thus, if we assume that $u$ depends on an additional variable $\tau$ then one may directly verify that the evolution equation
\begin{equation}
	u_\tau=u_{nnn}+\frac{3 (\anotherpropcoeffsquared+\xisquared)}{2}\arbfuningauss^2 u_n-\frac{\xisquared}{2} u_n^3
	\label{symmetr}
\end{equation}
constitutes a higher symmetry of the Gauss equation (\ref{mtzgen}). The latter is indeed compatible with the above third-order equation. In Case (A), (\ref{symmetr}) becomes the mKdV equation
\begin{equation}
 u_\tau = u_{nnn} + \frac{3}{2}u^2u_n
\end{equation}
so that the compatibility condition $u_{s\tau}=u_{\tau s}$ related to (\ref{normalform})$_1$ produces the potential mKdV (pmKdV) equation
\begin{equation}
 \rho_\tau = \rho_{nnn} + \frac{1}{2}\rho_n^3,
\end{equation}
which is known to be compatible with the sine-Gordon equation (\ref{sine-Gordon}) \cite{rogersschief}.

In Case (C), we obtain the modified mKdV (m$^2$KdV) equation \cite{calogerodegasperis,fokas}
\begin{equation}\label{mmkdv}
 u_\tau = u_{nnn} + \frac{3}{2}u_n\sin^2u + \frac{1}{2}u_n^3
\end{equation}
which generates the $\tau$-evolution
\begin{equation}\label{evolrho}
 \rho_\tau = u_{nn}\cos u + \frac{1}{2}u_n^2\sin^2u + \frac{1}{2}u_n^3
\end{equation}
via compatibility. By construction, the latter is compatible with (\ref{backlund})$_2$. Moreover, the evolution equations (\ref{mmkdv}) and  
(\ref{evolrho}) imply that
\begin{equation}\label{mkdv2}
  {(u\pm\rho)}_\tau = {(u\pm\rho)}_{nnn} + \frac{1}{2}{(u\pm\rho)}_n^3
\end{equation}
if one takes into account (\ref{backlund})$_2$. This is consistent with the compatibility of the pmKdV equations (\ref{mkdv2}) with the sine-Gordon equations (\ref{sine}).

\subsection{Tzitz\'eica connections}

Case (3) is characterised by the constraint
\begin{equation}\label{tziconst}
 \psi'' = -\frac{\kappa}{2}(\psi' - \kappa\psi),\quad \kappa = \pm\sqrt{-K},\quad K <0.
\end{equation}
Standard (computer algebra) algorithms (see, e.g., \cite{adler}) may be used to verify the generalised symmetries
\begin{equation}
	u_\xi=\frac{u_s^2 u_{nss}-u_s u_{ns} u_{ss}}{\arbfuningauss}+\frac{2}{3 \pctz}\frac{u_s u_{ns}^3}{\arbfuningauss^3}-\frac12 \frac{(\pctz\arbfuningauss+2\arbfuningauss')u_{ns} u_s^3}{\arbfuningauss^2}
\label{anothersymmtz}
\end{equation}
and
\begin{equation}
\eqalign{
	u_\tau=u_{nnnnn}-\frac52 \pctz\,u_{nnn}\left(\frac{\pctz}{2} u_n^2-u_{nn}+\arbfuningauss(\pctz\arbfuningauss+\arbfuningauss')\right)\\
\phantom{u_\tau=u_{nnnnn}}-\frac52\pctz\, u_n u_{nn}\left(\frac{\pctz}{2} u_{nn}+(\pctz\arbfuningauss+\arbfuningauss')^2\right)\\
\phantom{u_\tau=u_{nnnnn}}+ u_n \left(\frac{\pctz^4}{16}u_n^4+\frac54\pctz^2\arbfuningauss^2(\pctz\arbfuningauss+\arbfuningauss')^2\right).}
	\label{mtzsym5}
\end{equation}
This assertion may be made good by verifying compatibility with the Gauss equation (\ref{mtzgen}). Remarkably, the stationary reduction $u_\xi=0$ of (\ref{anothersymmtz}) produces a compatible third-order equation of the type (\ref{compateq}) with 
\begin{equation}
	F = \frac{\pctz}{2} \frac{u_{ns} u_s}{\arbfuningauss}+\frac{u_{ns} u_{ss}}{\arbfuningauss u_s}-\frac{2}{3\pctz}\frac{u_{ns}^3}{u_s\arbfuningauss^3},
	\label{FTzitz}
\end{equation}
while (\ref{mtzsym5}) may be found in a classification of integrable fifth-order evolution equations~\cite{mikhshab}. A connection with the integrable Tzitz\'eica equation of affine differential geometry is obtained by applying the Miura-type transformation
\begin{equation}
 3e^w = (\arbfuningauss'+\pctz\arbfuningauss) \left(\frac{\uvar_{ns}}{\arbfuningauss}\pm\frac{\sqrt{3}}{2}\pctz \uvar_s\right)
\label{mtzitztotzitz}
\end{equation}
which provides a link to the third-order equation
\begin{equation}
w_{nns} = w_n(3 e^w-2 w_{ns}).
	\label{tz3oeq}
\end{equation}
In fact, the latter is independent of the constant $\kappa$ so that one may regard $\kappa$ as a spectral parameter for this third-order equation as in the original Miura transformation \cite{Miura1968,AblowitzClarkson1991} for the KdV equation. Thus, if $u$ is a solution of the Gauss equation (\ref{mtzgen}) with $\psi$ being a solution of (\ref{tziconst}) then $w$ as defined by (\ref{mtzitztotzitz}) is a solution of (\ref{tz3oeq}).

The differential equation (\ref{tz3oeq}) may be integrated once to obtain the Tzitz\'eica equation \cite{Tzitzeica0708,rogersschief} in the form
\begin{equation}
w_{ns}=e^w+h(s)e^{-2 w}.
\end{equation}
Its standard form given by $h(s)=\pm1$ may be obtained by applying suitable transformations of the type $s\rightarrow S_1(s)$ and $w\rightarrow w + S_2(s)$. Hence, we may regard the Gauss equation considered in this subsection as a ``modified Tzitz\'eica equation'' in the terminology of soliton theory. Its Lax pair 
\begin{equation}\label{lax}
\Psi_n=U\Psi,\quad\Psi_s=V\Psi
\end{equation}
may be obtained by applying a gauge transformation to the standard Lax pair of the Tzitz\'eica equation (see, e.g., \cite{rogersschief}). One obtains the matrix-valued functions
\begin{equation}
\eqalign{
U= 
\left(\begin{array}{ccc}
0 &\quad \psi'+\kappa\psi\quad & \psi'+\kappa\psi \\[1mm]
\lambda(\psi'+\kappa\psi) & 0 & \frac{\sqrt{3}}{3}(\psi'+\kappa\psi) \\[1mm]
-\lambda(\psi'+\kappa\psi) & \frac{\sqrt{3}}{2}\kappa\psi & 0 \\
\end{array}
\right)\\
V=
\left(\begin{array}{ccc}
0 & \frac{\sqrt{3}}{6}\frac\kappa\lambda u_s & -\frac{1}{3\lambda}\frac{u_{ns}}{\psi} \\[2mm]
\frac{1}{3}\frac{u_{ns}}{\psi} & \frac{\kappa}{2}u_s-\frac{\sqrt{3}}{3}\frac{u_{ns}}{\psi} & -\frac{\sqrt{3}}{3}\frac{u_{ns}}{\psi}\\[2mm]
\frac{\sqrt{3}}{6}\kappa u_s  &\quad -\kappa u_s+\frac{\sqrt{3}}{3}\frac{u_{ns}}{\psi}\quad & -\frac{\kappa}{2}u_s+\frac{\sqrt{3}}{3}\frac{u_{ns}}{\psi}
\end{array}
\right),}
\end{equation}
wherein $\lambda$ is the spectral parameter. One may directly verify that the linear system (\ref{lax}) for the vector-valued function $\Psi$ is indeed compatible if $u$ is a solution of the Gauss equation subject to (\ref{tziconst}).

As pointed out in the introduction, the above analysis has led to natural non-classical orthogonal coordinate systems on surfaces of constant Gaussian curvature. In particular, the discovery of orthogonal coordinate systems on pseudospherical surfaces which are related to the modified Tzitz\'eica equation is somewhat surprising since the underlying Lie algebraic structure of the Lax pairs associated with surfaces of constant non-vanishing Gaussian curvature is the Kac-Moody algebra $A_1^{(1)}$, while the graded loop algebra $A_2^{(2)}$ is usually associated with the Tzitz\'eica equation (see \cite{rogersschief} and references therein). It is also observed that (apparently different) modified Tzitz\'eica equations have been recorded elsewhere in the literature. For instance, a modified Tzitz\'eica equation and the associated Miura-type transformation containing a particular Weierstrass $\wp$-function have been presented in \cite{BorisovZykovPavlov02}.

\section{Determination of the fibres and fluid flow}
\label{posvector}
So far, we have treated the surface $M^2$ as a Riemannian manifold \cite{Eisenhart49} without considering its embedding in Euclidean space $\R^3$ which is determined by the second fundamental form \cite{Eisenhart60}
\begin{equation}
\mbox{II} = e\, ds^2 + 2 f\, ds dn + g\, dn^2,
\end{equation}
the coefficients of which obey the Mainardi-Codazzi equations
\begin{equation}
\eqalign{
\mce_n-\mcf_s=\mce\frac{\uvar_{ns}}{\uvar_s}+\mcf\left(\frac{\arbfuningauss'\uvar_s}{\arbfuningauss}-\frac{\uvar_{ss}}{\uvar_s}\right)+\mcg\frac{\uvar_s\uvar_{ns}}{\arbfuningauss^2} \\
\mcf_n-\mcg_s=-\mce\frac{\arbfuningauss\arbfuningauss'}{\uvar_s}+\mcf\left(\frac{\arbfuningauss'\uvar_n}{\arbfuningauss}-\frac{\uvar_{ns}}{\uvar_s}\right)-\mcg\frac{\arbfuningauss'\uvar_s}{\arbfuningauss}.}
	\label{MC}
\end{equation}
In the case of a surface of constant Gaussian curvature $K$, this system is completed by adding the defining expression for the Gaussian curvature \cite{Eisenhart60}
\begin{equation}\label{Gauss}
K=\frac{eg-f^2}{u_s^2\psi^2}.
\end{equation}
Thus, for a given solution $(u,\psi)$ of the third-order equation (\ref{mtzgen}), any solution $(e,f,g)$ of the system (\ref{MC}), (\ref{Gauss}) uniquely determines a surface $M^2$ parametrised by the position vector $\boldr(s,n)=(x(s,n),y(s,n),z(s,n))$ up to a Euclidean motion. The fibres are then represented by the curves $\boldr(s,n=\mbox{const})$ on $M^2$.

Conversely, if a parametrisation $\boldr(\sigma,\nu)$ of a particular surface $M^2$ of constant Gaussian curvature $K$ on which the motion takes place is given then it is required to determine the change of coordinates $\sigma=\sigma(s,n)$ and $\nu=\nu(s,n)$ which transforms the given metric 
\begin{equation}
\mbox{I} = \boldr_\sigma^2\,d\sigma^2 + 2\boldr_\sigma\cdot\boldr_\nu\, d\sigma d\nu + \boldr_\nu^2\,d\nu^2
\end{equation}
into $g$ as given by (\ref{metric}). In order to illustrate this procedure, we solve this problem in the case of the three simplest examples of surfaces of constant Gaussian curvature, namely the plane, the sphere and the pseudosphere.

\subsection{Motions on the plane}

The simplest surface of Gaussian curvature $K=0$ (developable surface) is the plane which we parametrise by
\begin{equation}
 x = \sigma,\quad y = \nu,\quad z=0
\end{equation}
without loss of generality. Hence, the coordinate transformation  $\sigma=\sigma(s,n)$, $\nu=\nu(s,n)$ is determined by
\begin{equation}\label{planexy}
  \sigma_s^2 + \nu_s^2 = u_s^2,\quad \sigma_s\sigma_n + \nu_s\nu_n = 0,\quad \sigma_n^2 + \nu_n^2 = \psi^2.
\end{equation}
The latter may be solved by algebraically parametrising the derivatives of $\sigma$ and $\nu$ in terms of an {\em a priori} arbitrary function $\phi(s,n)$ according to
\begin{equation}
	\sigma_s=u_s\cos\phi,\quad \sigma_n=-\psi \sin\phi,\quad \nu_s=u_s\sin\phi,\quad \nu_n=\psi\cos\phi.
	\label{posvectxy}
\end{equation}
However, the associated compatibility conditions $\sigma_{sn}=\sigma_{ns}$ and $\nu_{sn}=\nu_{ns}$ determine $\phi$ up to a constant of integration. Indeed, one obtains the pair
\begin{equation}
	\phi_s=-\frac{u_{ns}}{\psi}, \quad \phi_n=\psi',
	\label{flatphisys}
\end{equation}
which is compatible modulo the Gauss equation (\ref{mtzgen}) for $K=0$. It is observed that the relations (\ref{posvectxy}) and (\ref{flatphisys}) coincide with those found in \cite{schiefquart,schieftmf} for steady planar motions of fibre-reinforced fluids.

\subsection{Motions on the sphere}

The unit sphere is the simplest surface of Gaussian curvature $K=1$. If we use spherical polar coordinates
\begin{equation}
	x= \sin \sigma\cos \nu,\quad y=\sin \sigma \sin \nu,\quad z=\cos \sigma
\end{equation}
to parametrise the unit sphere then the re-parametrisation $\sigma=\sigma(s,n)$, $\nu=\nu(s,n)$ yields the relations
\begin{equation}
	\sigma_s^2+\nu_s^2\sin^2\sigma = \uvar_s^2,\quad \sigma_s \sigma_n+\nu_s \nu_n\sin^2\sigma=0,\quad \sigma_n^2+\nu_n^2\sin^2 \sigma = \arbfuningauss^2.
	\label{pqsn}
\end{equation}
As in the flat case, it is convenient to introduce a new quantity $\phi$, leading to the system
\begin{equation}\label{sigmanu}
\eqalign{
	 \sigma_s = \uvar_s\sin \phi,\quad  \sigma_n = \psi\cos\phi,\\
       \nu_s = \uvar_s\cos \phi \csc \sigma,\quad \nu_n = -\psi \sin\phi \csc \sigma,}
\end{equation}
the compatibility of which results in the pair
\begin{equation}
	\phi_s= \uvar_s\cos \phi \cot \sigma-\frac{\uvar_{ns}}{\psi},\quad\phi_n= \psi'-\psi \sin\phi \cot \sigma.
	\label{compatsys}
\end{equation}
It is observed that the system (\ref{sigmanu})$_{1,2}$, (\ref{compatsys}) is decoupled from (\ref{sigmanu})$_{3,4}$ which may be integrated once $\sigma$ and $\phi$ are known. One may also readily verify that the pair (\ref{compatsys}) is compatible modulo the Gauss equation (\ref{mtzgen}) for $K=1$. 

The vector fields underlying the Frobenius system (\ref{sigmanu})$_{1,2}$, (\ref{compatsys}) may be shown to generate the $so(3)$ Lie algebra. In fact, this observation gives rise to the linearising transformation
\begin{equation}\label{defphi}
  \phi = \arctan \frac{\phi_2}{\phi_3},\quad \sigma = \arctan\left[\frac{\phi_3}{\phi_1}\sqrt{1 + \frac{\phi_2^2}{\phi_3^2}}\,\right],\quad \bphi = \left(\begin{array}{c} \phi_1\\ \phi_2\\ \phi_3\end{array}\right),
\end{equation}
where $\bphi$ obeys the linear system
\begin{equation}\label{linearsphere}
  \bphi_s = \left(\begin{array}{ccc} 0 & -u_s & 0\\ u_s & 0 & -u_{ns}/\psi\\ 0 & u_{ns}/\psi & 0\end{array}\right)\bphi,\quad
  \bphi_n = \left(\begin{array}{ccc} 0 & 0 & -\psi\\ 0 & 0 & \psi'\\ \psi & -\psi' & 0\end{array}\right)\bphi.
\end{equation}
It is observed that the above linear system admits the first integral
\begin{equation}\label{first}
  \phi_1^2 + \phi_2^2 + \phi_3^2 = \hat{\kappa}^2 = \mbox{const}.
\end{equation}
Thus, any solution $(\phi_1,\phi_2,\phi_3)$ of the linear system (\ref{linearsphere}) is mapped to a solution $(\phi,\sigma)$ of the nonlinear system (\ref{sigmanu})$_{1,2}$, (\ref{compatsys}). Conversely, for any given solution of the nonlinear system (\ref{sigmanu})$_{1,2}$, (\ref{compatsys}), the functions $\phi_1$, $\phi_2$ and $\phi_3$ defined by (\ref{defphi})$_{1,2}$ and (\ref{first}) obey the linear system (\ref{linearsphere}). Finally, in terms of $\bphi$, the remaining quantity $\nu$ is given by
\begin{equation}
  \nu = \hat{\kappa}\int\mbox{sgn}(\phi_1)\left(\frac{\phi_3}{\phi_2^2+\phi_3^2}u_s\,ds - \frac{\phi_2}{\phi_2^2+\phi_3^2}\psi\,dn\right).
\end{equation}

\subsection{Motion on the pseudosphere}

The classical pseudosphere is the simplest surface of Gaussian curvature $K=-1$. It admits the parametrisation \cite{rogersschief}
\begin{equation}
	x=\sin \sigma\cos \nu,\quad y=\sin \sigma \sin \nu,\quad z=\cos \sigma+\ln\left(\tan\frac{\sigma}{2}\right)
\end{equation}
with $0<\sigma<\pi$, in which case the connection between the two sets of coordinates is given by
\begin{equation}
\eqalign{
	\sigma_s^2\cot^2 \sigma +\nu_s^2 \sin^2 \sigma=\uvar_s^2,\quad \sigma_s \sigma_n \cot^2 \sigma+\nu_s \nu_n \sin^2 \sigma=0\\ \sigma_n^2\cot^2\sigma +\nu_n^2 \sin^2\sigma=\psi^2.}
\end{equation}
The compatibility of the associated parametrisation
\begin{equation}
\eqalign{
\sigma_s=\uvar_s\sin\phi\tan \sigma,\quad \sigma_n=\psi\cos\phi\tan \sigma\\ \nu_s=\uvar_s\cos\phi\csc \sigma,\quad \nu_n=-\psi\sin\phi\csc \sigma}	
\label{pseudopq}
\end{equation}
leads to the pair
\begin{equation}\label{phipsi}
	\phi_s = u_s\cos\phi -\frac{u_{ns}}{\psi},\quad \phi_n = \psi'-\psi\sin \phi
\end{equation}
which is, once again, compatible modulo the Gauss equation (\ref{mtzgen}) for $K=-1$. In the current case, once the differential equations (\ref{phipsi}) have been solved, the remaining functions $\sigma$ and $\nu$ may be found iteratively via quadratures. 

It is observed that the structure of the system (\ref{phipsi}) implies that it may be transformed into a system of Riccati equations and, hence, be linearised. In fact, it turns out that (\ref{pseudopq})$_{1,2}$ plays an important role in this connection. Thus, if we consider, for instance, the upper half of the pseudosphere represented by $\pi/2\leq \sigma <\pi$ then one may directly verify that the change of variables
\begin{equation}
\phi = 2\arctan\frac{\phi_1}{\phi_2},\quad \sigma = \pi - \arcsin (\phi_1^2+\phi_2^2),\quad \bphi = \left(\begin{array}{c} \phi_1\\ \phi_2\end{array}\right)
\end{equation}
linearises the coupled system (\ref{pseudopq})$_{1,2}$, (\ref{phipsi}) to obtain
\begin{equation}
 \bphi_s = \frac{1}{2}\left(\begin{array}{cc} 0 & u_s-u_{ns}/\psi\\ u_s + u_{ns}/\psi & 0 \end{array}\right)\bphi,\quad
 \bphi_n = \frac{1}{2}\left(\begin{array}{cc} -\psi & \psi'\\ -\psi' & \psi \end{array}\right)\bphi
\end{equation}
and the remaining function $\nu$ is given by
\begin{equation}
  \nu = - \int\left(\frac{\phi_1^2-\phi_2^2}{{(\phi_1^2+\phi_2^2)}^2}u_s\,ds + \frac{2\phi_1\phi_2}{{(\phi_1^2+\phi_2^2)}^2}\psi\, dn\right).
\end{equation}

\section*{Conclusion}


We have demonstrated that large classes of steady motions of ideal
fibre-reinforced fluids on surfaces of constant Gaussian curvature are
governed by various integrable specialisations of the Gauss equation
which constitutes a third-order partial differential equation.
Integrable systems theory may now be applied to construct explicitly
such motions. On the one hand, one can apply B\"acklund transformations \cite{rogersschief}
to generate sequences of fibre distributions from a seed distribution
for a fixed surface $M^2$ of constant Gaussian curvature. On the other
hand, one can apply B\"acklund transformations to generate sequences of
surfaces $M^2$ on which the motion takes place. The concrete application
of those two types of B\"acklund transformations will be discussed
elsewhere. It would be interesting to investigate whether the generation
of those two types of sequences commutes, that is, whether the order in
which the sequences of motions for a given surface and the sequences of
surfaces for a given motion are generated is of importance. In this
connection, it is observed that the commutativity of B\"acklund
transformations is an important topic in soliton theory \cite{rogersschief}.

\section*{References}


\end{document}